\documentclass[superscriptaddress,amsmath,amssymb,aps,epsf,prl,twocolumn]{revtex4-2}
\usepackage[T1]{fontenc}
\usepackage{newunicodechar}
\usepackage{graphicx}
\usepackage{dcolumn}
\usepackage{bm}
\usepackage{amsmath}
\usepackage{amsfonts}
\usepackage{amssymb}
\usepackage[dvipsnames]{xcolor}
\usepackage{pinlabel}
\usepackage{ragged2e}

\graphicspath{{Figures/}}

\begin{document}

	\title{Strange metal behavior from incoherent carriers scattered by local moments.}

	\author{Sergio Ciuchi}
	\affiliation{Dipartimento di Scienze Fisiche e Chimiche, Universit\'a dell’Aquila, Coppito-L’Aquila, Italy and Istituto dei Sistemi Complessi, CNR, 00185 Roma, Italy}

	\author{Simone Fratini}
	\affiliation{Universit\'e Grenoble Alpes, CNRS, Grenoble INP, Institut N\'eel, 38000 Grenoble, France}

\begin{abstract}
We study metallic transport in an effective model that describes the coupling of electrons to fluctuating 
magnetic moments  with full SU(2) symmetry, exhibiting characteristic behavior of metals at the approach of the Mott transition. 
We show that scattering by fluctuating local moments
causes a fully incoherent regime of electron transport with  T-linear resistivity. This strange metal regime is characterized by  almost universal, nearly Planckian slope and a finite zero-temperature intercept, that we can associate respectively to the amplitude fluctuations and to the random orientations of local magnetic moments. Our results indicate a route for understanding the microscopic origin of strange metal behavior that is unrelated to quantum criticality and does not rely on the existence of quasiparticles.
\end{abstract}

\maketitle

\noindent

\paragraph{Introduction.---} 
One of the active theoretical challenges in quantum materials is the explanation of strange metal behavior: in many 
complex metals the resistivity increases approximately linearly with temperature over an extended temperature range, as $\rho(T) \simeq \rho_0 + B T$ instead of the expected quadratic behavior, therefore contradicting the very foundations of Fermi liquid theory. 
The slope of $\rho(T)$ has been the subject of thorough investigation, and is now often interpreted as an indication of a fundamental "Planckian" bound for inelastic scattering \cite{Zaanen,Hartnoll-Mackenzie,Bruin,Legros,Phillipsscience2022}. 
The constant term $\rho_0$ has received less attention; the fact that both in theory \cite{Terletska11,Vucicevic-vtx} and in the experiments  \cite{Kanoda1} this constant contribution shifts rigidly the resistivity curves up and down upon tuning the interaction strength is strongly reminiscent of the effects of elastic disorder \cite{Kanoda2}.  The precise origin of such intrinsically generated disorder is however unclear.

In this work we study high temperature transport in a microscopic model of electrons coupled to fluctuating local moments, that captures the typical behavior of metals at the approach of the Mott transition. This setup enables a transparent understanding of the microscopic transport processes in terms of the statistical properties of magnetic scatterers. In agreement with standard treatments of correlated electron transport in the framework of the  Hubbard model \cite{Kokalj,Devereaux,PakiraPRB2015,Perepelitsky,Tremblay-PNAS22,Gull-slopeinvariance-PRR20} we find that scattering by magnetic fluctuations leads to a resistivity that varies linearly with temperature with a slope that is qualitatively in agreement with Planckian transport. Importantly, we identify the random orientations of SU(2) magnetic moments, whose size is controlled by the strength of the electron-electron interactions,  as being at the origin of the elastic residual resistivity $\rho_0$.  We generalize the present treatment to include  scattering effects in the charge sector, showing that the addition of even modest disorder can stabilize linear resistivity down to the lowest temperatures, as is observed in a variety of correlated metals.

\paragraph{Model and method.---}
We consider an interacting model where  the electron spins are coupled to bosonic degrees of freedom, mimicking the interaction of electrons with fluctuating magnetic moments. In the framework of single-site Dynamical Mean Field Theory (DMFT) this is encoded in the following local impurity action $S=S_0+S_{int}+S_{b}$ with:
\begin{eqnarray}
S_0 &=& \int d\tau \int d\tau^\prime \sum_\sigma \bar{c}_\sigma(\tau) G_0^{-1}(\tau-\tau^\prime) c_\sigma(\tau^\prime) ,
\label{eq:actionS0}
\\
S_{int}&=&-\sum_{\sigma,\rho} \sum^3_{\nu=0} g_\nu \int d\tau \bar{c}_{\sigma}(\tau) \boldsymbol{\sigma}^{\nu}_{\sigma,\rho} c_{\rho}(\tau) X^\nu,
\label{eq:actionSint}
\\
S_{bos}&=&+\frac{\beta}{2}\sum_\nu k_\nu (X^\nu)^2.
\label{eq:actionSb}
\end{eqnarray}
$S_0$ describes the Weiss field resulting from the integration of the lattice electrons  \cite{DMFTReview}, that we take to be independent on spin indices $\sigma$ since we do not consider symmetry broken phases.   We consider a semi-elliptical density of states (DOS) 
with half-bandwidth $D$ and set the particle density to half filling. 

The interaction term, $S_{int}$, 
describes electrons (Grassman variables $c_\sigma(\tau),\bar{c}_\sigma(\tau)$) moving in the thermalized  environment generated by  classical bosonic variables $X^\nu$, governed by the harmonic term $S_{bos}$. 
The vector part ($\nu = 1,2,3$) describes the interaction of electrons with fluctuating local moments.
Formally, a SU(2) symmetric spin-boson interaction can be rigorously derived by linearizing the Hubbard model via a Hubbard-Stratonovich transformation and then taking the high temperature limit, yielding $g_{\nu}=1$ and $k_{\nu}=3/U$ ($\nu=1,2,3$) \cite{ScheurerPNAS2018}.  The scalar part of $S_{int}$ (index $\nu = 0$) nominally represents a local coupling with the charge, that has been thoroughly studied elsewhere \cite{Millis}. In a broader sense, the latter can also serve as a proxy for retarded, longer-ranged Coulomb interactions between electrons.
All along this work we consider an isotropic spin coupling $g_{1,2,3}=g_s$ preserving SU(2) symmetry; in this case the relevant quantity that couples to the electron spin is the energy fluctuation associated to the radial bosonic variable $v=g_s r=g_s \sqrt{\sum^3_{\nu=1} (X^\nu)^2}$, with $\lambda_s=g^2_s/2 k_s D$ the corresponding dimensionless coupling strength   (see \cite{Tiwari-EPL14} for a real-space implementation of the theory, and \cite{Okamoto-Millis} for a related DMFT approach for Ising spins). Connecting to the original Hubbard model via the Hubbard-Stratonovich derivation yields $(U/D)=6\lambda_s$.

Because we are interested in high temperature transport, we take the classical limit for the bosons.
The validity of this approximation is quite general in regimes where the relevant bosonic (spin and/or charge) fluctuations are slow, a situation that is commonly realized in the presence of strong electronic correlations.
The local Green's function $G$ can then be calculated as a statistical average over the  distribution ${\cal P}(v)$ of local boson energy fluctuations (details in SM). The corresponding electron self-energy $\Sigma$ is  obtained via the usual self-consistency condition, $\Sigma(\omega)=\omega-\frac{D^2}{4} G(\omega)-G^{-1}(\omega)$ for the chosen semi-circular DOS.
The electrical conductivity is calculated via the Kubo formula:
\begin{equation}
\sigma(T) = 2\bar{\sigma}\pi \int d\epsilon \Phi(\epsilon) \int d\omega A^2(\epsilon,\omega)
\left(-\frac{d f}{d\omega}\right),
\label{eq:condmain}
\end{equation}
where $A(\epsilon,\omega)=-Im [\omega-\epsilon -\Sigma(\omega)]^{-1}/\pi$ is the spectral function, $f$ is the Fermi function and $\Phi(\epsilon)$ is the transport function of the lattice model, given by the DOS of the squared velocity along a given direction. 
Conductivity units are set by $\bar{\sigma}=e^2a^2/\hbar \Omega$ with $a$ the lattice parameter in the relevant direction and $\Omega$ the unit cell volume (cf. SM, Sec. II). 

\paragraph{Correlated behavior and Mott transition.---}
\begin{figure}[t!]
\includegraphics[width=1.0\columnwidth]{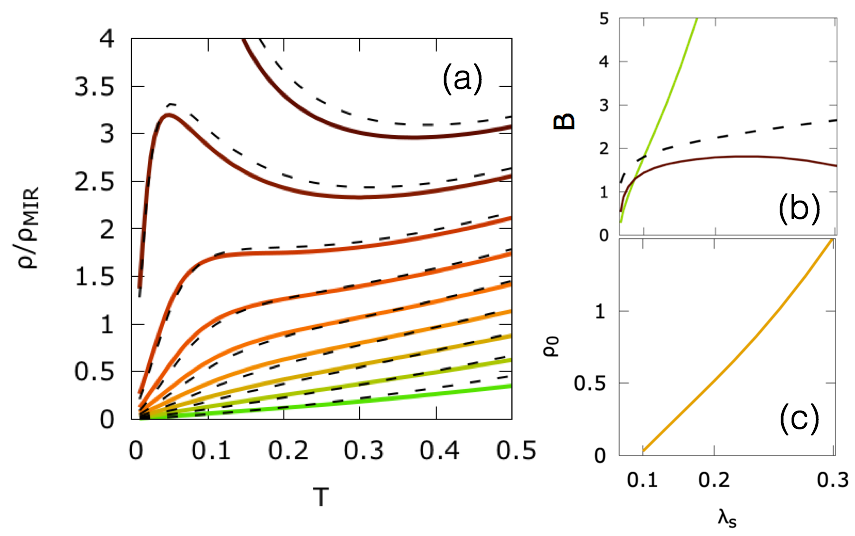}
\centering
\caption{(a) Temperature dependent resistivity calculated for equally spaced $\lambda_s=0.04 \to 0.36$, in units of $\rho_{MIR}$. 
Dashed lines are the results of the incoherent approximation  (see text).
(b) Temperature coefficient of the resistivity $B$  at low ($T=0.05$, green) and high temperature ($T=0.5$, brown) as a function of $\lambda_s$ in quadratic scale, together with the prediction of the incoherent approximation (dashed). (c) The zero temperature intercept $\rho_0$ (residual resistivity) extrapolated from high temperatures.}
\label{fig:allresistivities}
\end{figure}

Fig. \ref{fig:allresistivities} shows resistivity vs temperature curves for increasing values of the coupling parameter $\lambda_s$. The results  reproduce the behavior observed at the approach and across the Mott transition in the Hubbard model \cite{Terletska11,Vucicevic-vtx}.
The metal-insulator transition (MIT) is found here at  $\lambda_s^c=0.36$,  corresponding to  $(U/D)_c=2.2$ in quantitative agreement with the single-site CTQMC result $(U/D)_c=2.3$ \cite{Toschi-MIT}. 

The present treatment correctly addresses the behavior of the resistivity at temperatures higher than the Fermi liquid temperature, where we find two qualitatively different linear regimes.
At low temperatures, only states near the Fermi energy contribute to transport. In this "resilient quasiparticle" regime \cite{resilient}, one can tentatively apply the weak coupling, low-T limit of Eq.(\ref{eq:condmain}), namely $\rho/\rho_{MIR} = \Gamma/2D$. Observing that the coupling to thermal bosons  
yields  $\Gamma \propto \lambda_s T $ \cite{MillisA15} leads to
a resistivity that is trivially linear in temperature, with a large variability in slopes upon varying $\lambda_s$ and a common  intercept $\rho_0=0$. This seems to  agree with what is seen in Fig. \ref{fig:allresistivities} (a) at low T. The slopes shown in Fig. \ref{fig:allresistivities} (b) (green), however, markedly contradict this weak-coupling prediction, as they increase with $\lambda_s^2$ \cite{Kiely-PRB21} instead of $\lambda_s$. We shall come back to this contradiction in the next paragraphs. 

\begin{figure}[t!]
\includegraphics[width=1.0\columnwidth]{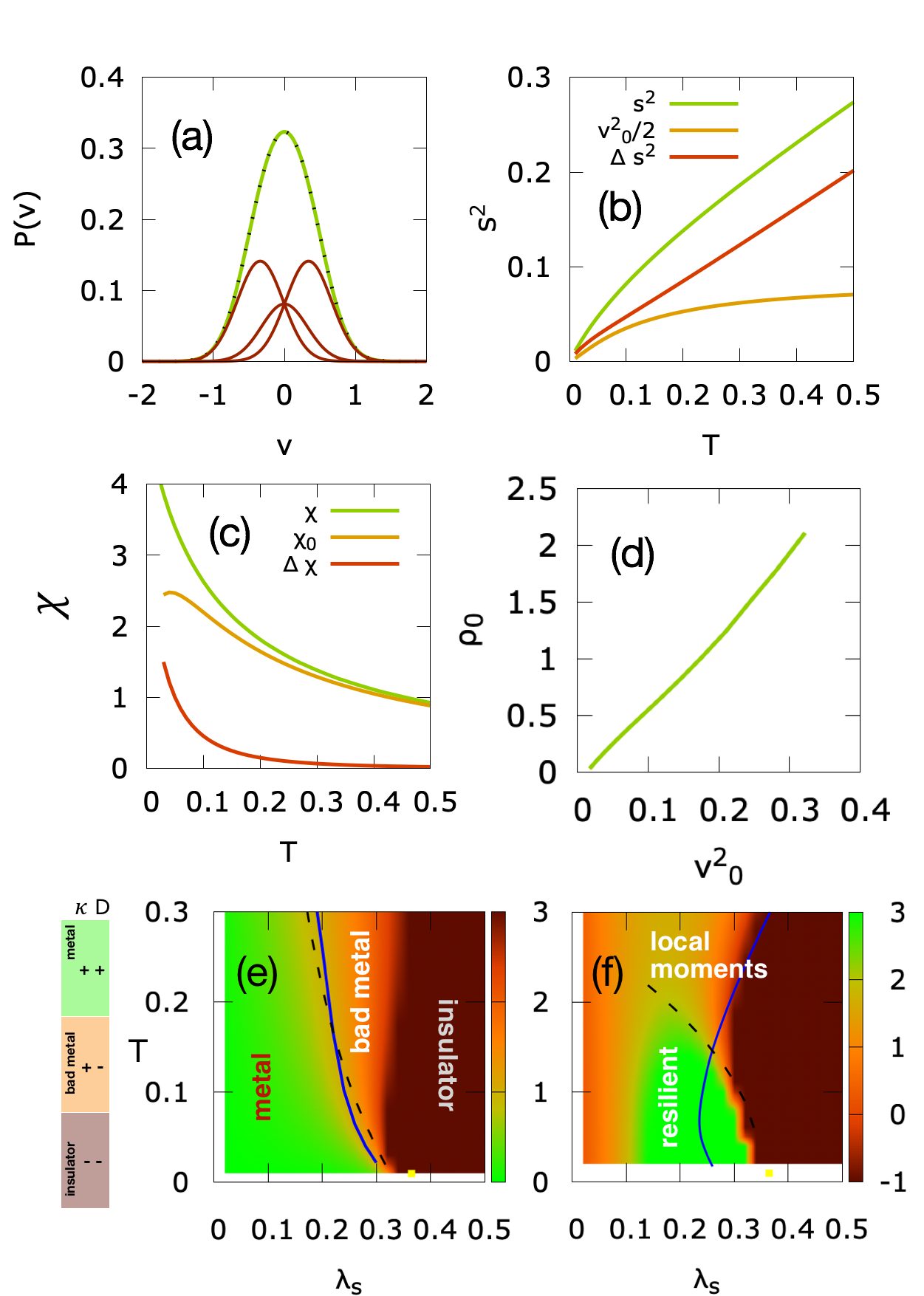}
\centering
\caption{ (a) Symmetrized distributions $P(v)$  of the SU(2) bosonic variable for $T=0.25$ and $\lambda_s=0.20$, decomposed into the singly occupied, spin up (right Gaussian), spin down (left) and empty/doubly occupied sites (center).  (b) Mean square fluctuation $s^2=\langle v^2 \rangle/3$  as a function of temperature for $\lambda_s=0.20$ (green), compared with the residual term $v^2_0$ (yellow) and the fluctuations $\Delta s^2$ (red) extracted from the multicomponent fits of $P(v)$. (c) Local spin susceptibility (green) compared with its analogous decomposition into $\chi_0$ (yellow) and $\Delta \chi$ (red). (d) Residual resistivity $\rho_0$ showing almost perfect correlation with the amplitude of the local moments $v^2_0$ obtained at high temperature. (e) Map of the resistivity and (f) of the slope $B$  in the $(\lambda_s,T)$ plane. In (e) the black dashed curve is the MIR limit $\rho=\rho_{MIR}$, the blue curve locates the existence of a well formed pseudogap in the spectrum, the yellow square is the location of the MIT (estimated at $T=0.01$). The left legend indicates the magnitude of the charge compressibility $\kappa$ and diffusivity $\cal{D}$ entering the Nernst-Einstein relation ($+/-$ stand for values $>/<1$, see SM). In (f) the  black dashed curve locates the formation of local moments, the blue curve is the point where the  boson distribution becomes bimodal. }
\label{fig:moments}
\end{figure}

For sufficiently large $\lambda_s$ still within the metallic phase  the system enters a second linear regime upon increasing the temperature:
here the slope $B$ is similar for all curves, while the $\rho_0$ intercept is nonzero and is strongly parameter-dependent
(Fig. \ref{fig:allresistivities}(a) and (c)). The slope itself is of order of unity when expressed in the natural units of the model, i.e. $B=d(\rho/\rho_{MIR})/d(T/D) \simeq 1-2$ (Fig. \ref{fig:allresistivities}(b) and map Fig. \ref{fig:moments}(f)), in quantitative agreement with theoretical studies on the Hubbard model \cite{Kokalj,Devereaux,PakiraPRB2015,Perepelitsky,Tremblay-PNAS22,Gull-slopeinvariance-PRR20} as well as with recent transport measurements of correlated organic metals as a function of pressure \cite{Kanoda1}. 
If we were to interpret this result within the weak scattering/low temperature Drude picture, as is customarily done in experiments \cite{Bruin,Legros,Delacretaz17,Hartnoll-Mackenzie}, we would obtain $d\Gamma/dk_BT=2 B$, yielding $d\Gamma/dT$ of order of few $k_B$, with a clear Planckian flavor.   The overall behavior found in this high-temperature regime is very reminiscent of the phenomenology observed in strange metals.
 
\paragraph{Emergence of local moments.---}
We now show that the strange metal behavior found at high T is a direct consequence of scattering from fluctuating local moments. 
Fig. \ref{fig:moments}(a) shows a typical radial distribution of the bosonic spin variable $v$, in the high temperature metallic regime at moderate $\lambda_s=0.2$ $(U/D=1.2)$. While at first sight the distribution seems Gaussian as predicted by perturbation theory (see SM), closer inspection reveals a hidden internal structure.
The distribution obtained numerically  can be perfectly described as the sum of three Gaussians of center $\pm v_0,0$ and variance $\Delta s$, originating respectively from singly occupied and empty/doubly occupied sites; 
their mean and variance represent the average magnitude  and  amplitude fluctuations of the corresponding magnetic moments.

Fig. \ref{fig:moments}(b) shows the numerical results for the total  mean square fluctuation $s^2=\langle v^2\rangle/3$ as well as the evolution of the fitting parameters $\Delta s^2$ and  $v_0^2/2$ with temperature, representing respectively the  amplitude and orientational fluctuations ($s^2=\Delta s^2 + v^2_0/2$ in the  high temperature regime of relevance here).  The amplitude fluctuations closely follow the classical thermal dependence, $\Delta s^2\simeq 2\lambda_s T$. More interesting to us, $v_0$ clearly shows the emergence of local moments; these are precursors of the Mott phase attained at large $\lambda_s$.  
The resistivity of electrons scattered by randomly oriented moments is readily evaluated to 
$\rho_0\propto v_0^2$.  This behavior is fully compatible with the numerical data of Fig.  \ref{fig:allresistivities}(c), confirming that the orientational fluctuations of the local moments are at the origin of the zero temperature intercept of the resistivity.

Finally, from the knowledge of $s^2$  we can derive the {\it local} spin susceptibility \cite{Sangiovanni2006} shown in 
Fig. \ref{fig:moments}(c). The latter appears to be dominated by the preformed moments, exhibiting  the familiar Curie  form $\chi_s\sim 1/T$ even at high temperatures where the thermal fluctuation part $\Delta s$ is much larger than the mean $v_0$.

\paragraph{Transport phase diagram.---}
The relation between the statistical properties of the spin variable and the transport mechanism is  illustrated in Fig. \ref{fig:moments}(d-e), showing  maps of the resistivity and its derivative in the $(\lambda_s,T)$ plane. 
Fig. \ref{fig:moments}(e) shows a progressive evolution of the resistivity as a function of the interaction strength $\lambda_s$, with values steadily increasing  from the metal (left) to the Mott insulator (right). Bad metal behavior occurs in the intermediate coupling regime, where the resistivity rises above the MIR limit (blue line). The bad metal regime defined by this condition is seen to coincide with  the region where the momentum-integrated spectral density $A(\omega)=-Im G(\omega)/\pi$ shows a  well-formed pseudogap \cite{Tiwari-EPL14} (dashed line, $A(0)$ equal to  half of its maximum value, see also SM). 

Fig. \ref{fig:moments}(f) shows a map of the slope $B$ in the same parameter range. Confirming the insights gained in the preceding paragraphs, we see large variations of the slope upon varying $\lambda_s$ at low temperatures. At high T, however, the slope is essentially constant in a very broad region of parameters, that we associate with the fluctuating local moments phase. This is delimited by the dashed line, defined as the temperature at which the amplitude $v_0$ of the local moments  stabilizes above  $80\%$ of its high-T saturation value: the color map shows that the stabilization of local moments coincides with the end of the resilient quasiparticle regime and the onset of strange metal transport, where the resistivity is T-linear with almost universal slope.
Also shown as a blue line is the line beyond which  the boson distribution becomes bimodal; this coincides with the change of sign of the resistivity slope at high temperature and merges into a  polaronic transition at $T=0$ (this is a precursor of the $T=0.0$ MIT, similar to the case of charge coupling only, see SM and \cite{Millis,Capone2003}). 

\paragraph{Origin of strange metal transport.---}
We now show that the character of the charge transport mechanism  in the high temperature regimes presented here is markedly non-Drude, i.e. it cannot be directly ascribed to  (not even ill-defined) quasiparticle properties; the resistivity is {\em not} proportional to the scattering rate. 
To this aim we make a drastic simplification and rewrite the Kubo formula Eq. (\ref{eq:condmain}) by  neglecting the band dispersion altogether. This corresponds to replacing the spectral function with its local (momentum integrated) equivalent $A(\omega)$, leading to $\sigma_{inc}  = 2 \pi \bar{\sigma} \int d\epsilon \Phi(\epsilon) \int d\omega [A(\omega)]^2 (-df/d\omega) $, 
with  $\int d\epsilon \Phi(\epsilon) =D^2/4$ on the Bethe lattice. The incoherent part of the resistivity, obtained through this formula, accurately describes the numerical data up to a constant that is readily determined by enforcing  $\rho(T=0)=0$ in the noninteracting limit.
This  is shown as black dashed curves in   Fig. \ref{fig:allresistivities}. Note that the incoherent formula applies down to the resilient quasiparticle regime, where standard weak-coupling approaches already break down (see Fig. \ref{fig:allresistivities}(c) and discussion).

With this paradigm shift at hand we can better understand the origin of the almost universal slope observed in the strange metal regime. Contrary to the low T limit, where the conductivity is  determined by the value of the spectral function at the Fermi level, at  high-T excitations throughout the entire electronic spectrum are involved in the transport process.  
The leading T-dependence  of the resistivity can  be evaluated by taking the high temperature limit of the Fermi function, $-df/d\omega \to 1/4T$, showing that as long as the temperature is comparable or larger than the range of variation of the spectral function $A(\omega)$, the  resistivity is linear in $T$ with a slope that  is controlled by the inverse of the spectral integral $I=\int d\omega [A(\omega)]^2$, namely  
$d(\rho/\rho_{MIR})/dT = B $ with $B \simeq (16/3\pi^2)/I$. The main effect of bosonic fluctuations is to mildly broaden the excitation spectrum, leading to an increase of the effective bandwidth, $D^*>D$ \cite{HoHu}. Evaluating the integral yields  $B\simeq D^*/D = 1 + O(\lambda_s T/D)$, which predicts that the resistivity slope is of order one to leading order in $\lambda_s T/D$. 
The dashed line in Fig. \ref{fig:allresistivities}(b) shows that the dimensionless slope for incoherent transport  stabilizes around $B\approx 2$ in the whole range where strange metal is observed, in good agreement with the full numerical results (full brown).
Because it follows from global properties of the electronic spectrum, this result is largely independent on the dimensionality, it does not rely  on the existence of a quantum critical point, 
and it is robust upon combining different sources of scattering, as we show  next.

\paragraph{Interplay with  disorder and phonon scattering.---}
\begin{figure}[ht!]
\includegraphics[width=0.9\columnwidth]{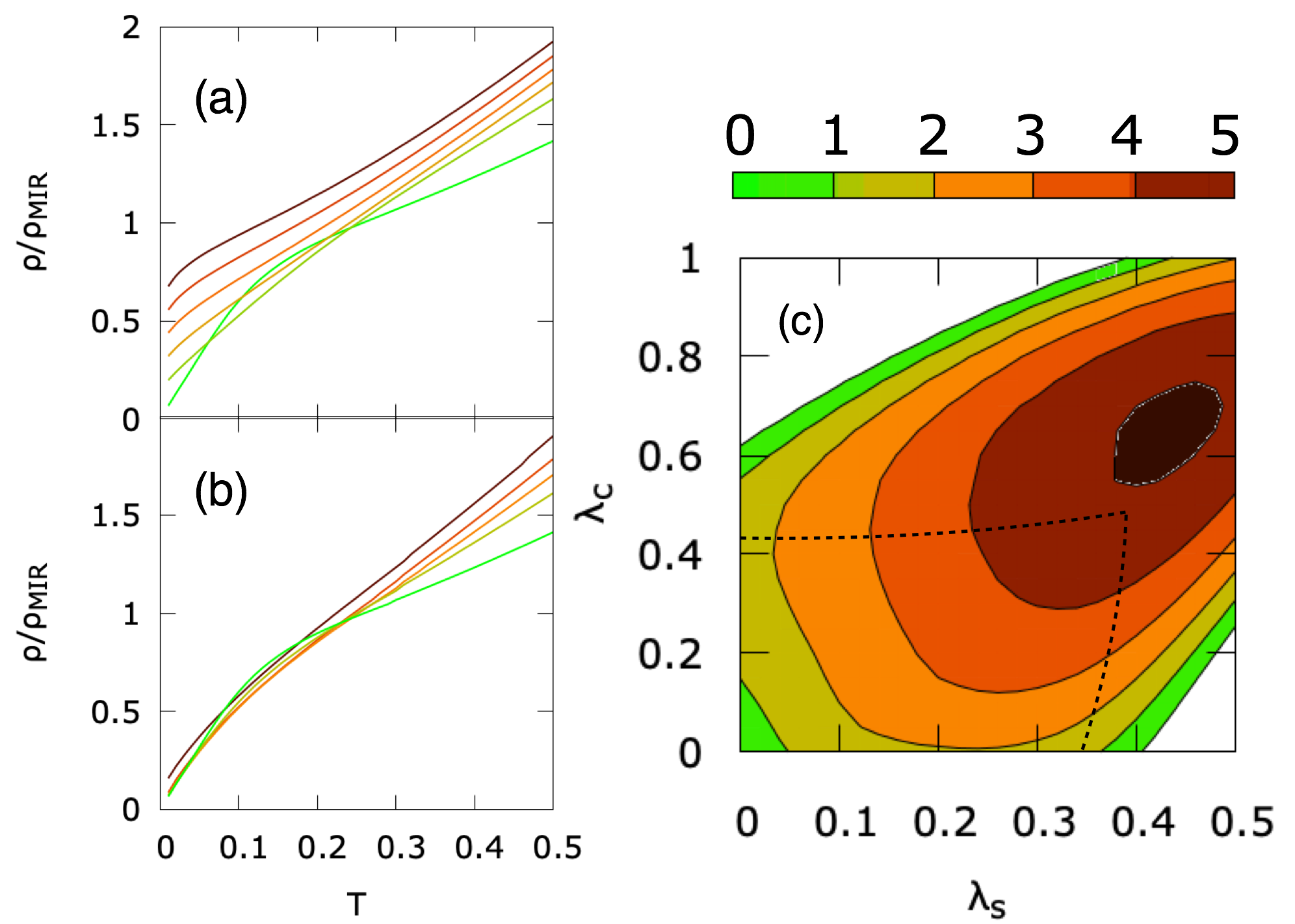}
\centering
\caption{ (a) Resistivity vs $T$ for $\lambda_s=0.2$ and different amounts of gaussian disorder with mean square fluctuation $\sigma^2=0.0 \to 0.5$  and (b)  different values of the charge coupling  $\lambda_c=0.00 \to 0.32 $ (from green to brown). (c) Map of the slope  at $T=0.5$ in the $(\lambda_s,\lambda_c)$ plane; the black dashed line marks the MIT  (estimated at $T=0.01$).}
\label{fig:disorderandcharge}
\end{figure}

Fig.  \ref{fig:disorderandcharge}(a) shows the effect of an added extrinsic disorder in the form of a random distribution of site energies (here taken to be Gaussian with standard deviation $\sigma$). Strikingly, addition of a modest amount of disorder  stabilizes strange metal behavior all the way down to the lowest temperatures. The different resistivity curves mainly differ by the value of their zero temperature intercept, with a  common slope that is largely unaffected by disorder. This agrees with what is observed in  transport experiments of correlated systems under irradiation \cite{Vobornik99,Rullier-Albenque03,Kanoda2}.

Fig. \ref{fig:disorderandcharge}(b) shows the resistivity calculated in the presence of an added coupling of a scalar boson to the electron charge $\lambda_c=g^2_0/2kD$,  for $\lambda_s=0.2$.  Also in this case the clear distinction between the resilient quasiparticle regime and the strange metal found at $\lambda_c=0$ is rapidly lost. This is mostly due to the fact that the charge coupling competes with the electronic correlations, shifting the Mott transition to higher values of  $\lambda_s$. This competition is most clearly seen in the phase diagram of Fig. \ref{fig:disorderandcharge}(c), showing  the locus of the MIT (dashed line) superimposed on the map of the high-temperature slope $B$. 

\paragraph{Concluding remarks.---}
The incoherent scattering of electrons from fluctuating local moments is able to explain the high-temperature strange metal behavior often observed in correlated electron systems. 
At odds with the normal Fermi liquid picture, in this regime the electrical conduction involves all states in the electronic spectrum instead of being governed by quasiparticle states near the Fermi energy alone: this leads to high-T linear resistivities with almost universal slopes that are insensitive to microscopic model details and qualitatively compatible with Planckian theoretical estimates. 
The microscopic mechanism unveiled here relies
entirely on local physics, and is therefore alternative to existing hydrodynamic approaches to strange metals
\cite{LucasHartnollPNAS2017,vucicevic23}. Interestingly, this "no quasiparticle" viewpoint helps understanding the origin of the mysterious compensation of the T-dependences of diffusivity and compressibility, seemingly conspiring to provide an overall $\rho \propto T$ behavior \cite{Kokalj,Mousatov,Kiely-PRB21}: in the strange metal regime the resistivity itself is the relevant physical quantity embodying the linear temperature dependence.

Remarkably, the inclusion of other sources of disorder in addition to the considered fluctuating local moments stabilizes  strange metal behavior down to the lowest temperatures,  as observed in a variety of correlated metals. While identifying the precise nature of this missing randomness goes beyond the scope of this work, the generality of the experimental observations would hint at an intrinsic origin. Nearly-frozen, self-generated disorder brought by frustrated (magnetic \cite{Frachet2020} or charge \cite{Mahmoudian,Mousatov,Driscoll,SciPost}) interactions and surviving down to the lowest temperatures could be a plausible candidate.

Finally, due to its modest computational cost the approach introduced here could serve as a meaningful starting point for realistic simulations of electronic correlations in materials, as well as in all these systems (interfaces, mesoscopic devices and disordered systems) where the effects of spatial inhomogeneity beyond single- and few-site clusters are crucial.

\acknowledgments
S.C. acknowledges funding from  NextGenerationEU National Innovation Ecosystem grant ECS00000041 - VITALITY - CUP E13C22001060006. 

\bibliography{sh.bib}

\end{document}